\documentstyle[prl,preprint,aps,psfig,12pt]{revtex}
\begin{document}
\draft
\preprint{IUCM98-006}
\title{Magnetic Anisotropy in Quantum Hall Ferromagnets}
\author{T. Jungwirth$^{1,2}$, S.P. Shukla$^{3}$, L. Smr\v{c}ka$^2$, 
M. Shayegan$^{3}$, and A.H. MacDonald$^{1}$}
\address{$^{1}$Department of Physics, Indiana University, Bloomington IN 47405}
\address{$^{2}$Institute of Physics ASCR,
Cukrovarnick\'a 10, 162 00 Praha 6, Czech Republic}
\address{$^{3}$Department of Electrical Engineering, Princeton University, 
Princeton NJ 08544} 
\date{\today}
\maketitle

{\tightenlines
\begin{abstract}

We show that the sign of magnetic anisotropy energy in quantum Hall
ferromagnets is determined by a competition between electrostatic and 
exchange energies.  Easy-axis ferromagnets tend 
to occur when Landau levels whose states have similar 
spatial profiles cross.  We report measurements of integer QHE evolution
with magnetic-field tilt.  Reentrant behavior observed for the 
$\nu = 4$ QHE at high tilt angles is attributed to easy-axis
anisotropy.  This interpretation is supported by a detailed calculation
of the magnetic anisotropy energy.

\end{abstract}
}


\narrowtext

In the quantum Hall effect (QHE) regime, two-dimensional electron
systems (2DES) can have ferromagnetic ground states in which 
electronic spins are completely aligned by an arbitrarily
weak Zeeman coupling\cite{dassarmabook}.  
However, spin-independence of the Coulombic
electron-electron interaction leads to isotropic Heisenberg 
ferromagnetism, and therefore to loss of ferromagnetic order
at any finite temperature\cite{2dreview}.  Richer physics occurs
when the two Landau levels that are nearly degenerate 
differ by more than a spin index.  For example,
double-layer QHE systems can be regarded as 
easy-plane (XY) two-dimensional ferromagnets\cite{ahmgregphil}
and exhibit a variety of effects
which have received considerable experimental\cite{dlexpt}
and theoretical\cite{dltheory} attention in recent years.
Idealized single-layer QHE systems have 
a phase transition\cite{giulianiquinn} in tilted magnetic fields 
between unpolarized and  spin-polarized states, and 
as we show below, can be regarded as easy-axis (Ising)
ferromagnets.  In this Letter we report experimental data for a
43~nm wide unbalanced GaAs quantum well in which a loss
of the QHE at $\nu = 4$  
is observed over a finite range of  magnetic-field tilt-angles. 
We derive a general expression for the magnetic 
anisotropy energy and propose that its sign is responsible for 
this observation.  We  show that 
in realistic quantum wells either easy-axis 
or easy-plane anisotropy can occur,
depending on spatial profiles of the 
orbitals of the crossing Landau levels.

We discuss the anisotropy energy first from a general
point of view, then specialize to 
two illustrative idealized examples before presenting realistic results
for the quantum well of interest.
In a strong magnetic field, the single-particle states 
of a 2DES are grouped into Landau levels with orbital 
degeneracy $N_{\phi} = A B / \Phi_0$, where $A$ is the 
system area, $B$ is the field strength, and $\Phi_0$ 
is the magnetic flux quantum.  We consider the case
where the Landau level filling factor $\nu \equiv N/N_{\phi}$
is an integer\cite{fraction}
and two different groups of $N_{\phi}$ orbitals are close to 
degeneracy.  We assume that other Landau levels are far 
enough from the Fermi energy to justify their neglect\cite{zheng}.
Using a {\em pseudospin} language\cite{ahmgregphil} 
to represent the Landau level index degree of freedom,
the class of Hamiltonians 
we consider can be expressed, up to an irrelevant constant, in the form 
\begin{eqnarray}
&H& = - b \sigma(\vec q=0) 
+ \frac{1}{2A} \sum_{\vec q} 
\left\{V_{\rho,\rho}(\vec q) \rho(-\vec q) \rho(\vec q) +
V_{\sigma,\sigma}(\vec q)\right.\times
\nonumber \\
&{ }&
\sigma(-\vec q) \sigma(\vec q)+
\left.V_{\rho,\sigma}(\vec q) \left[\rho(-\vec q) \sigma(\vec q)
+ \sigma(-\vec q) \rho(\vec q)\right] \right\}.
\label{hamiltonian}
\end{eqnarray} 
In Eq.~(\ref{hamiltonian}), $b$ is half the energy separation 
between the nearly degenerate Landau levels, and 
$\rho(\vec q)$ and $\sigma(\vec q)$
are respectively the sum and difference of the density  
operators\cite{leshouches} projected onto the up and down 
pseudospin  Landau levels.  
Note that $b$ is half the {\it single-particle} energy difference 
and does not include mean-field contributions 
from Coulomb or exchange interactions with electrons in the 
Landau levels of interest.  For simplicity, we 
have limited the present discussion
to cases for which the total number of electrons with each 
pseudospin index is conserved.  The effective interactions 
that appear in Eq.~(\ref{hamiltonian}) are related
to the effective interactions between pseudospins by the following relations:
$V_{\rho,\rho} = (V_{\uparrow,\uparrow} + V_{\downarrow,\downarrow}
+ 2 V_{\uparrow,\downarrow})/4$,
$V_{\sigma,\sigma} = (V_{\uparrow,\uparrow} + V_{\downarrow,\downarrow}
- 2 V_{\uparrow,\downarrow})/4$, and 
$V_{\rho,\sigma} = (V_{\uparrow,\uparrow} - V_{\downarrow,\downarrow})/4$.

Our calculation of the pseudospin anisotropy energy is based
on the following single Slater determinant wavefunction:
\begin{equation}
|\Psi[\hat n] \rangle = 
\prod_{m=1}^{N_{\phi}} c^{\dagger}_{m,\hat n} |0\rangle \; .
\label{wavefunction}
\end{equation} 
Here $m$ labels the orbital states within a Landau level and 
$\hat n$ denotes the pseudospinor aligned in the 
$\hat n = [\sin(\theta)\cos(\phi),\sin(\theta)\sin(\phi),\cos(\theta)]$
direction.  This many-particle state is fully pseudospin
polarized\cite{caveat1,future}.  A straightforward calculation yields 
the following result for the dependence of energy on pseudospin
orientation:
\begin{equation} 
\frac{\langle \Psi[\hat n] | H | \Psi[\hat n] \rangle }{N} =
- b^* \cos(\theta) + \frac{U_{\sigma,\sigma}}{2} \cos^2(\theta)\; . 
\label{anisoeng}
\end{equation}
Here $b^* = b - U_{\rho,\sigma}$ and for all indices 
\begin{equation}
U_{s,s'} = \int \frac{d \vec q}{(2 \pi)^2} [V_{s,s'}(\vec q=0)
- V_{s,s'}(\vec q)] \exp ( - q^2 \ell^2/2)\; ,
\label{anisointegral}
\end{equation}
where
$\ell=\sqrt{\hbar c/eB}$ is the magnetic length.
In Eq.~(\ref{anisoeng}) we have dropped terms in the energy that  
are independent of pseudospin orientation.  The 
right hand side of this equation is independent of $\phi$ 
because the $\hat z$ component of total pseudospin is a 
good quantum number.

For each effective field strength $b^*$, the pseudospin
orientation is determined by minimizing the total energy. 
For $U_{\sigma,\sigma} > 0 $, easy-plane anisotropy, $ \cos(\theta)
=0$ at $b^*=0$ and the pseudospin evolves continuously with
effective field as illustrated in Fig.~\ref{anis}(a), reaching alignment
for $|b^*| > U_{\sigma,\sigma}$.  For $U_{\sigma,\sigma} < 0$, 
easy-axis anisotropy, local minima occur
at both $\cos(\theta) = 1$ and $\cos(\theta) = -1$ for 
$|b^*|<|U_{\sigma,\sigma}|$.
If only global pseudospin rotation processes were possible,
macroscopic energy barriers would separate these two locally
stable states, resulting in hysteretic behavior (see 
Fig.~\ref{anis}(a)).  The sign of $U_{\sigma,\sigma}$ is 
determined by competition between the 
two terms in square brackets on the right hand side 
of Eq.~(\ref{anisointegral}).  The $V_{\sigma,\sigma}(\vec q=0)$
term is an electrostatic energy which
is present when the two pseudospin states have different charge density
profiles perpendicular to the electron layers.  This term favors
easy-plane anisotropy..  The $V_{\sigma,\sigma}(\vec q)$ term is  
the exchange energy which favors easy-axis anisotropy  
which will always occur when $V_{\sigma,\sigma}(\vec q)$
is an increasing function of wavevector.

Transport measurements in the QHE regime are 
extremely sensitive to the energy gap for charged excitations.
Generally, large energy gaps give rise to well developed Hall
plateaus and deep minima in the dissipative resistivity.
In the Hartree-Fock approximation, the 
quasiparticle energy gap of anisotropic QHE ferromagnets
can be written quite generally as\cite{tomasthesis}
\begin{equation}
\label{gap}
\Delta_{HF}=I_{\uparrow\downarrow} -2 U_{\sigma\sigma} +
\frac{2 b^*}{\cos(\theta)}\; ,
\end{equation} 
where $I_{\uparrow\downarrow}=\int\frac{dq^2}{(2\pi)^2}\,
\exp\left(-q^2\ell^2/2\right)\left(V_{\rho\rho}-V_{\sigma\sigma}\right)$.
For the easy-plane case, $\Delta_{HF}$ is a continuous
function of the effective field $b^*$, decreasing linearly for 
$b^*/U_{\sigma\sigma} < -1$, constant for $|b^*|/U_{\sigma\sigma} < 1$
and increasing linearly for $b^*/U_{\sigma\sigma} > 1$.
In contrast, if the system has easy-axis
anisotropy, $\Delta_{HF}$ decreases to $I_{\uparrow\downarrow}$
at the extremes 
of the hysteresis loop ($b^*/U_{\sigma\sigma} =  \pm 1$) 
before jumping to $I_{\uparrow\downarrow} + 4|U_{\sigma\sigma}|$ 
when the pseudospin magnetization reverses. 
In Fig.~\ref{anis}(b) we summarize the above results  
by plotting the renormalized Hartree-Fock gap $\Delta^{*}_{HF}=(\Delta_{HF}-
I_{\uparrow\downarrow})/2|U_{\sigma\sigma}|$ as a function of
$b^*/|U_{\sigma\sigma}|$.   In the Hartree-Fock approximation
this quantity depends only on the sign of the anisotropy energy. 

For concreteness, we  mention two idealized models
which we regard as paradigms for the easy-plane and easy-axis anisotropy
cases.  For two arbitrarily narrow quantum
wells separated by a distance $d$ with full polarization of
the true electron spin, we let 
pseudospin represent the layer index\cite{fullpolarized}.
The ``pseudospin'' Zeeman field $b$ is then 
proportional to the bias electric field, $E_g$, created by a gate 
external to the electron system: $ b = e E_g d/2 $.
On the other hand, for a single arbitrarily narrow quantum
well with $\nu = 2 m$ in which the real-spin Zeeman
coupling has been increased\cite{giulianiquinn} so as 
to bring the up-spin $n=m$ Landau level close to 
degeneracy with the down-spin $n=m-1$ Landau level, we let  
the pseudospin represent the spin-index of the Landau level close
to the Fermi energy.  The pseudospin Zeeman coupling for this model is 
$ b = (g^{*} \mu_B B - \hbar \omega_c +
I_0)/2$. Here the first term is the real-spin
Zeeman coupling, the second term is the cyclotron energy and the last term 
is the contribution to $b$ from exchange interactions with 
frozen Landau levels lying well below the Fermi energy ($I_0/(\sqrt{\pi/2}
\, e^2/\epsilon\ell)$ = 1/2, 5/16, and 31/128 for $m$=0, 1, and 
2 respectively\cite{giulianiquinn,mcdojiliu}).
The effective Coulomb interaction energies for the two 
models are summarized in Tab.~\ref{tab}.
For the ideal double-layer 
model, the electrostatic term $V_{\sigma,\sigma}(q=0)$ dominates,
$V_{\sigma,\sigma}(q)$ is a monotonically 
decreasing function of $q$ and $U_{\sigma,\sigma}$ is  
positive.  On the other hand for the ideal tilted-field 
model, the pseudospin wavefunctions differ only
in the plane of the 2DES, the electrostatic term is consequently
absent, and the exchange term produces easy-axis 
anisotropy  
($U_{\sigma,\sigma}/(\sqrt{\pi/2}
\, e^2/\epsilon\ell)$ = -3/16, -33/256, and -107/1024 for $m$=0, 1, and 
2 respectively).

Now we turn to the discussion of the measured QHE
evolution with tilted field, shown in 
Fig.~\ref{rxx}\cite{fractions}.
In finite width quantum wells, 
the large tilt angles necessary to bring the up and down pseudospin
Landau levels close to degeneracy
result in substantial coupling of the in-plane 
component of the magnetic field to orbital degrees of freedom\cite{tomas}.
These orbital effects can be incorporated  
\cite{future} by adjusting the effective 
interactions appropriately.  In particular, for real finite-width quantum
wells, the perpendicular charge density profiles of the two pseudospin Landau
levels differ, and the electrostatic contribution to 
$U_{\sigma,\sigma}$ is no longer zero.  The sign of
the anisotropy energy depends in detail on the quantum
well geometry, the tilt angle and the filling factor. The 
insets in Fig.~\ref{rxx}(b) show charge-density 
profiles in the studied quantum well 
for the relevant orbitals at high tilt angles obtained from 
self-consistent LDA calculations\cite{tomas}:
$n = 0, \downarrow$ and $n = 1, \uparrow$ 
at $\nu = 2$; $n = 1, \downarrow$ and $n = 2, \uparrow$ 
at $\nu = 4$.  From these orbitals we 
obtain\cite{future} that, for $\nu = 2$, $U_{\sigma,\sigma}$ 
increases substantially with tilt angle, is only marginally negative
for  $b^*=0$ which occurs at $\theta=72^o$ and
becomes positive at larger $\theta$ (see Fig.~\ref{rxx}(b). 
This results demonstrates that easy-plane anisotropy can 
occur in realistic single quantum wells. 
If so, referring to the quasiparticle gap predictions 
summarized in Fig.~\ref{anis}, 
a strong QHE may be expected throughout the  
region of tilt angles where the relevant Landau levels
are close to degeneracy.  
Consistent with this expectation, the experimental data of 
Fig.~\ref{rxx} show a strong minimum at $\nu=2$ at all angles near
$\theta=72^o$ and no clear evidence for the disappearance of the QHE
is  observed up to the highest accessible tilt-angles for $\nu=2$.
Like the weak dependence of quasiparticle gap on 
bias potential, noted in experimental studies of
double-quantum-well systems\cite{ezawa}, this robustness 
of the QHE is a general property of easy-plane QHE ferromagnets. 

At $\nu = 4$, our calculations predict 
that $b^*=0$ occurs at $\theta=79^o$, and that the 
density profiles of the two pseudospin states are similar even at 
high tilt angles,
as illustrated in Fig.~\ref{rxx}. Hence, 
$U_{\sigma,\sigma}$ is only weakly angle dependent and is still 
{\em negative} around
$\theta=79^o$.  We attribute the clear degradation
of the measured  QHE   at $\nu =4$ to 
easy-axis anisotropy. The tilt angle $\theta=80^o$ where
the $\nu =4$ QHE disappears is in a good quantitative agreement with the
theoretically predicted  angle $\theta=79^o$ at which
the pseudospin Zeeman field
$b^*$ vanishes. 
We expect transport properties inside the hysteresis loop 
in the easy-axis case, to have a complicated disorder dependence.  
Spatially random potentials
couple differently to different Landau levels and 
will produce a random pseudospin magnetic field. 
This is expected\cite{imry} to lead to the formation of large
domains with particular pseudospin orientations.  
The dynamics of pseudospin reorientation is likely to be
controlled by barriers to domain wall motion.
If these are comparable to $k_B T$, the pseudospin will achieve 
alignment with the effective field on laboratory time scales,
$\cos(\theta)$ will change from $-1$ to $1$ at $b^* = 0$,
and the energy gap will have a cusp.
This scenario appears to apply for recent experiments which 
study analogous Landau level crossings in the valence band of 
GaAs\cite{pepper} and to some tilted field driven transitions at 
fractional Landau level filling factors\cite{fractions}.
On the other hand, when some domain wall motion barriers are much larger than
$k_B T$, we expect that all physical properties will exhibit hysteretic
behavior, and that the electronic state will have domain 
structure for $b^*$ close to zero.  Dissipation due to 
mobile charges created in domain walls\cite{future} can
then lead to a breakdown of the QHE.
We expect that dissipative and Hall resistances will then 
depend on measuring current and sample history,
as well as on temperature.

In the disorder free limit, easy-axis anisotropy in two-dimensions 
leads to a finite temperature continuous phase transition in 
the Ising universality class and stronger thermodynamic 
anomalies than for the Kosterlitz-Thouless phase transition 
of easy-plane systems.  The transition temperature can
be estimated\cite{future} by balancing energy and entropy terms in the 
free-energy of long domain walls:
\begin{equation}
k_B T_c  \sim U_{\sigma,\sigma} (w R / \ell^2)\; ,
\label{tceq}
\end{equation}
where $w$ is the domain wall width and $R$ is the 
domain wall orientation correlation length.  The domain 
wall physics of these easy-axis ferromagnets 
is unconventional because the spin-stiffness 
is negative\cite{future}.   Preliminary results from work presently
in progress suggest that $w R/ \ell^2$ is substantially larger
than one and that the critical temperature should 
typically exceed $\sim 1$ Kelvin.

{\em Note added}: A recent experimental study\cite{woowon}
we learned of after this work was completed
finds hysteretic behavior in a narrow (25~nm) GaAs quantum well
in vicinity of $\nu=2/5$ and 4/9 fractional QHE's
which correspond to integer QHE's at composite fermion filling
factors $\nu=2$ and 4 respectively. In these experiments, Zeeman
coupling strength was controlled both by 
applying hydrostatic pressure and by tilting the field.
We believe that the theory developed
in this paper explains the origin of the hysteresis
found in Ref.~\cite{woowon} at very low temperatures ($T
\stackrel{{\protect\textstyle <}}{\sim}$ 200mK). 
We have not observed similar hysteresis in our data (Fig.~\ref{rxx}); this may
be because of our higher available base temperature (300mK).

This work was supported by the National Science
Foundation under grants DMR-9623511, DMR-9714055 and INT-9602140, 
by the Ministry of Education of the Czech Republic under
grant ME-104 and by the Grant Agency of the Czech Republic
under grant 202/98/0085.

\begin{table}
\begin{center}
\begin{tabular}{cccc}
Model  &  $V_{\rho\rho}$ & $V_{\sigma\sigma}$ & 
$V_{\rho\sigma}$   
\\
\hline
double-layer & $(1+e^{-qd})/2q$ & $(1-e^{-qd})/2q$ & 0 
\\
tilted-field & $\frac{\left(L_m(q^2/2)+L_{m-1}(q^2/2)\right)^2}{4q}$ & 
$\frac{\left(L_m(q^2/2)-L_{m-1}(q^2/2)\right)^2}{4q}$ & 
$\frac{\left(L_m(q^2/2)\right)^2-\left(L_{m-1}(q^2/2)\right)^2}{4q}$ 
\\
 \end{tabular}
 \end{center}
 \caption{\protect
 Effective Coulomb interactions  in units of
 $2 \pi e^2\ell/\epsilon$ as a function of wavevector $q$ in units
 of $\ell^{-1}$ for  ideal double-layer and tilted-field
 models. $L_n(x)$ is the Laquerre polynomial.}
 \label{tab}
 \end{table}

\begin{figure}
\caption{Pseudospin orientation (a) and Hartree-Fock
quasiparticle gap (b) as a function of the effective field $b^*/
|U_{\sigma\sigma}|$ for the easy-axis (full line) and easy-plane
(dashed line) broken symmetry states.
}
\label{anis}
\end{figure}

\begin{figure}
\caption{(a):  Measured longitudinal resistance 
vs perpendicular component of the magnetic field 
for different tilt angles ($\theta$ is measured from the normal to 
the plane of the 2DES) at $T=300$~mK. 
The $\nu=4$ QHE is lost at $\theta\approx
80^o$ and reappears at $\theta\approx 80.5^o$ while for $\nu=2$ no loss
of the QHE is observed near $\theta=72^o$.
The inset shows calculated charge distribution at $\theta = 0$ for 
the unbalanced GaAs quantum well studied here.  The front-gate and
back-gate voltages and the 2DES
density ($N=1.57\times 10^{11} {\rm cm}^{-2}$) 
were fixed during the experiment. (b): Anisotropy 
energies calculated for the geometry of this 
sample at $\nu=2$ and 4. 
The two Landau levels which are 
brought close to degeneracy by applying
in-plane component of the magnetic field are indicated in the insets 
together with calculated density profiles for up (dashed line)
and down (solid line) pseudospin orbitals
at high tilt angles.
}
\label{rxx}
\end{figure}


\begin{references}
\bibitem{fullpolarized}  In this idealized model we assume
that the true spin-degree of freedom has been frozen by
the external magnetic field.
\end{references}
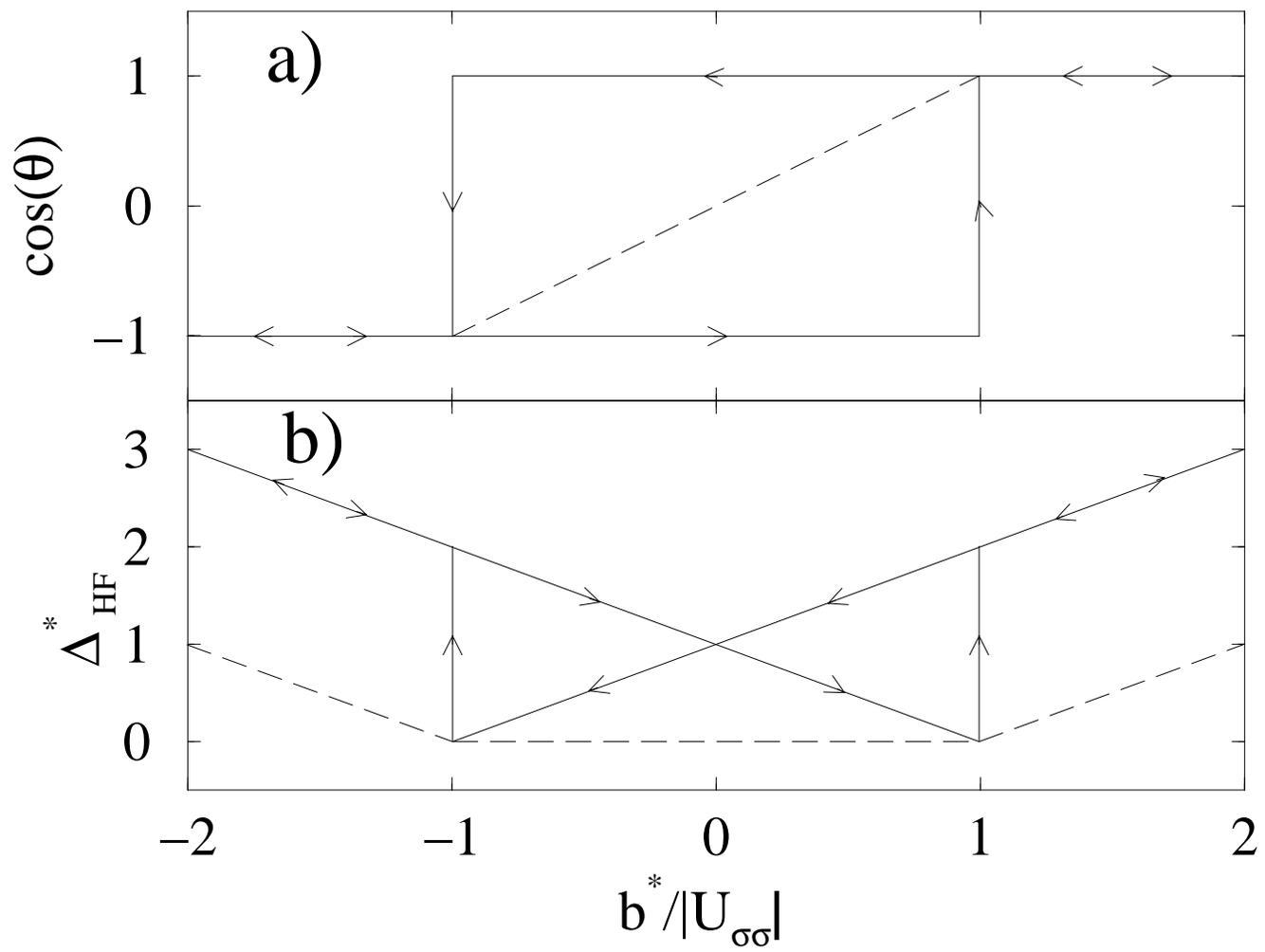
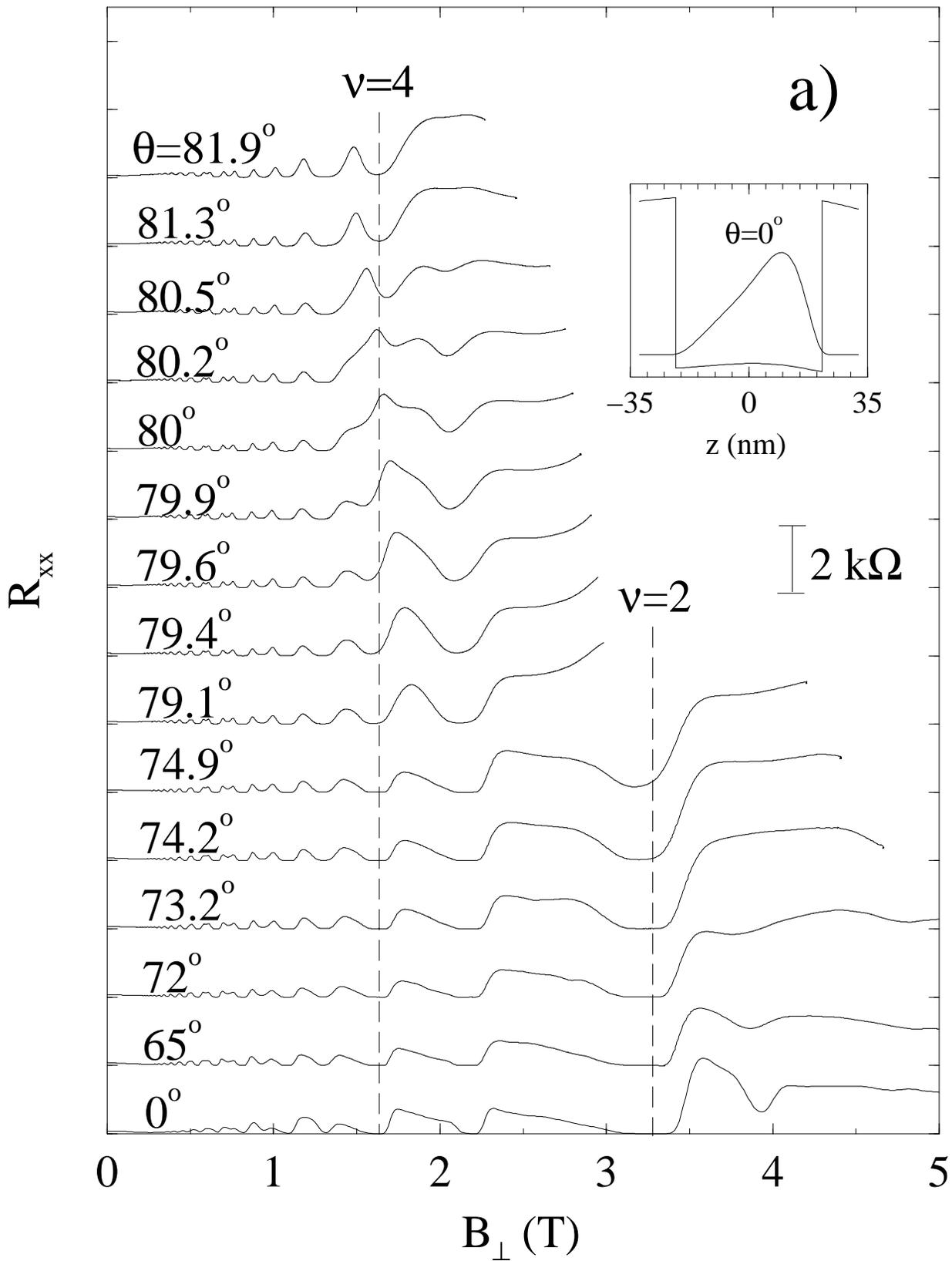
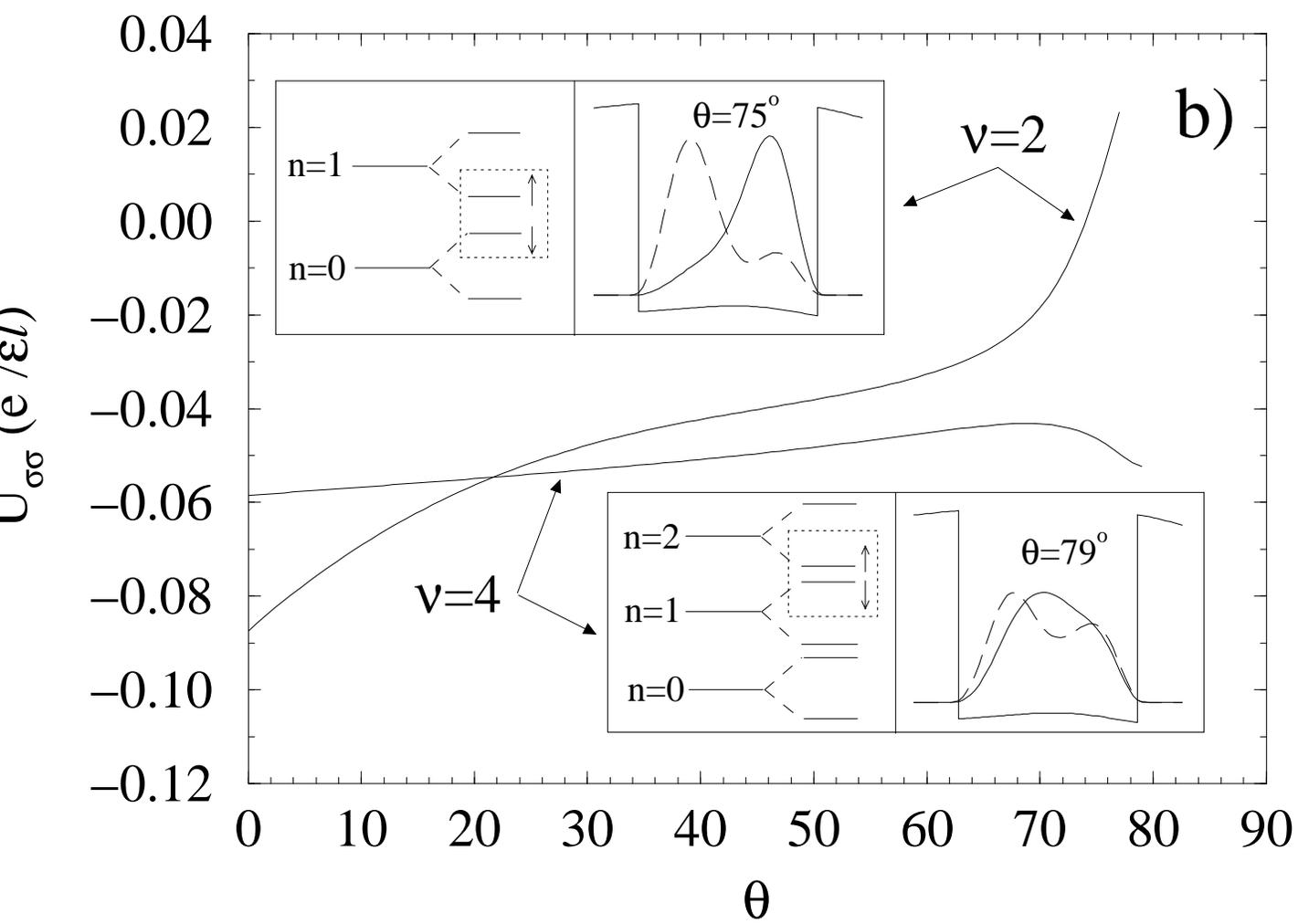
\end{document}